# Electrophoretic Mobilities of Dissolved Polyelectrolyte Charging Agent and Suspended Noncolloidal Titanium during Electrophoretic Deposition

Kok-Tee Lau[1,2] and C.C. Sorrell[1]

[1]School of Materials Science and Engineering

The University of New South Wales

Sydney, NSW 2052

Australia

[2]Faculty of Manufacturing Engineering

Universiti Teknikal Malaysia Melaka

76109 Durian Tunggal, Melaka

Malaysia

Contact: C.C. Sorrell

School of Materials Sci. & Eng.

University of New South Wales

Sydney, NSW 2052

Australia

Tel. (+61-2) 9385-4421

Fax (+61-2) 9385-5956

E-Mail C.Sorrell@unsw.edu.au

## Abstract

Coarse (≤20 μm) titanium particles were deposited on low-carbon steel substrates by cathodic electrophoretic deposition (EPD) with ethanol as suspension medium and poly(diallyldimethylammonium chloride) (PDADMAC) as polymeric charging agent. Preliminary data on the electrophoretic mobilities and electrical conductivities on the suspensions of these soft particles as well as the solutions themselves as a function of PDADMAC level were used as the basis for the investigation of the EPD parameters in terms of the deposition yield as a function of five experimental parameters: (a) PDADMAC addition level, (b) solids loading, (c) deposition time, (d) applied voltage, and (e) electrode separation. These data were supported by particle sizing by laser diffraction and deposit surface morphology by scanning electron microscopy (SEM). The preceding data demonstrated that Ti particles of ~1-12 μm size, electrosterically modified by the PDADMAC charging agent, acted effectively as colloidal particles during EPD.

Owing to the non-colloidal nature of the particles and the stabilization of the Ti particles by electrosteric forces, the relevance of the zeta potential is questionable, so the more fundamental parameter of electrophoretic mobility was used. A key finding from the present work is the importance of assessing the electrophoretic mobilities of *both* the suspensions and solutions since the latter, which normally is overlooked, plays a critical role in the ability to interpret the results meaningfully. Further, algebraic uncoupling of these data plus determination of the deposit yield as a function of charging agent addition allow discrimination between the three main mechanistic stages of the electrokinetics of the process, which are: (1) surface saturation; (2) compression of the diffuse layer, growth of polymer-rich layer, and/or competition between the mobility of Ti and PDADMAC; and (3) little or no decrease in electrophoretic mobility of Ti, establishment of polymer-rich layer, and/or dominance of the mobility of the PDADMAC over that of Ti.

## 1. Introduction

There are many conventional commercial methods for the achievement of surface hardening of steel, including electroplating, electrogalvanising, conversion coating, hot dip coating, metal cladding, porcelain enameling, fusion hardfacing, thermal spraying, vapor deposited coating, and surface hardening through heat treatment (such as pack cementation) [1]. Each of these methods has advantages and disadvantages in terms of applicability, ease of process, cost, and other issues. An alternative coating method that rarely has been considered for surface hardening is electrophoretic deposition (EPD). Although the literature on the EPD of *metallic* coatings is not extensive, the EPD of *ceramic* coatings has been studied many times in considerable detail. The interest in this method lies largely its advantages over other coating methods [2,3]:

- Potential to produce coatings of variable thickness (thin to thick film range)
- Potential for precision production of highly reproducible coatings in terms of microstructure and thickness
- Potential to apply even coatings on substrates of complex shapes
- Rapid deposition rates (seconds to minutes)

- Simplicity of process, requiring only simple equipment (power supply only)
- Low cost of infrastructure and process

The EPD process is similar to that of electroplating in that it is performed using only a d.c. power supply with cathode and anode immersed in a liquid-filled container. In the case of electroplating, the liquid is an ionic solution and dense metal is deposited while, in the case of EPD, the liquid is a suspension of colloidal particles (≤1 μm) and the porous deposit consists of these particles [2]. The applied electric field drives the charged particles toward the oppositely charged electrode, typically a cathodic substrate, on which they are deposited. In addition to the coating of conducting substrates, EPD also has been used to fabricate monolithic, laminated, and functionally graded free-standing objects as well as to infiltrate porous materials and woven fibre performs used in composite production.

Most of the applications and studies of EPD have used ceramic particles but there are a few publications on the EPD of particulate non-noble metals [4-15]. It is clear that one of the reasons for the lack of availability of such colloidal metal particles is the tendency for most of the metallic particles to oxidise, thereby forming a passivating oxide layer of a high volume ratio relative to the remaining metal core, which makes it an unattractive method to achieve a uniform metallic coating. On the other hand, successful EPD of noble metals, including gold, silver, and palladium, and their potential applications in the fabrication of electronic devices, have been explored [2,4,13]. However, the usage of these noble metallic particles is limited to high-end applications due to their high costs compared to those of base metals.

It is clear that coarse non-noble metallic particles have the advantage of lower volume ratios of surface oxide layers. The disadvantage of such particles is that they are non-colloidal and so have relatively low surface charge [15]. Hence, suspension in liquids is difficult owing to the reduced electrostatic attraction and the consequent deleterious effect of gravity. Lower surface charges and greater particle weight also decrease the mobilities of particles during electrophoretic deposition.

The shortcomings of insufficient surface charge and excessive weight potentially can be overcome through the use of polymeric charging agents, where the associated ionic groups provide additional surface charge and the polymeric chains provide steric stabilization. Two well known examples are the polyelectrolytes poly(diallyldimethylammonium chloride) (PDADMAC) [13,16] and polyethyleneimine (PEI) [13,15], which contain the ionic groups ammonium and imine, respectively. Further, these polymeric charging agents play an important role as binders to improve adhesion between deposited particles and substrate [2, 16]. The attachment of such charged polymers to particles and the resultant electrosteric forces between particles result in what are known as *soft particles* [17].

The aim of the present work was to examine the factors affecting the room-temperature EPD of relatively coarse titanium particles on low-carbon steel substrates using absolute ethanol as suspension medium and a PDADMAC polymeric charging agent. The interpretation of the EPD of soft particles does not appear to have been reported previously. Further, the critical role of the electrophoretic mobility of the solution appears to be unrecognized in studies of suspensions. The variables studied were addition level of PDADMAC, solids loading, deposition time, applied voltage and electrode separation. The parameters assessed were electrophoretic mobility, electrical conductivity, deposit yield, and surface morphology.

The present study is motivated by the potential for the controlled introduction of a uniform surface layer of metallic titanium particles on steel for the purpose of surface hardening of low-carbon steel by one of two potential routes:

- *Ex situ* hard coating: Surface hardening by titanium deposition and (a) subsequent nitridation or (b) graphite deposition and subsequent carburisation during heat treatment
- *In situ* diffusion coating: Surface hardening by titanium deposition, diffusion of titanium into steel during heat treatment, and concurrent carburisation.

In contrast to methods such as thin-film application (*ex situ*) and pack cementation (*in situ*), some of the advantages of the above two processes potentially are:

- More controllable process
- Less waste of raw materials
- Less expensive infrastructure and process
- More even coating on irregular shapes

In the present work, titanium particles were selected for EPD as a preliminary stage of surface hardening of steel owing to the potential for subsequent heat treatment to effect surface diffusion, nitridation, or carburization of the metallic titanium. Ethanol was selected as the dispersion medium owing to its non-corrosive behaviour (in comparison to water) and low cost. PDADMAC was selected as the charging agent owing to the retention of its strong cationic charge under a wide range of pH conditions [18].

## 2. Experimental Procedure

**Suspension:** A representative image of the as-received raw material used in the present work is given in Figure 1. The morphology of this titanium (Ti) powder (99.7 wt%, SE-Jong Materials Co. Ltd., South Korea) was platy, subangular, and of medium sphericity. The particle size range of this raw material is given in Figure 2, which shows that its range was ~1-50 μm, with a median size ($d_{50}$) of ~17 μm. Each suspension was made by adding 0.1 g of Ti powder to 20 mL of absolute ethanol (99.7 wt%, CSR Ltd., Australia) to give a solids loading of 5 mg/mL. The suspension was magnetically stirred at a speed of 400 rpm for 1 min using a 2 cm length Teflon-coated bar in a 25 mL Pyrex beaker. The polyelectrolyte poly(diallyldimethylammonium chloride) solution (PDADMAC, reagent grade, 20 wt% in water, average molecular weight 100,000-200,000, true density 1.04 g/mL, Sigma-Aldrich Co.) was added by pipette, followed by magnetic stirring for 30 min at the same stirring speed.

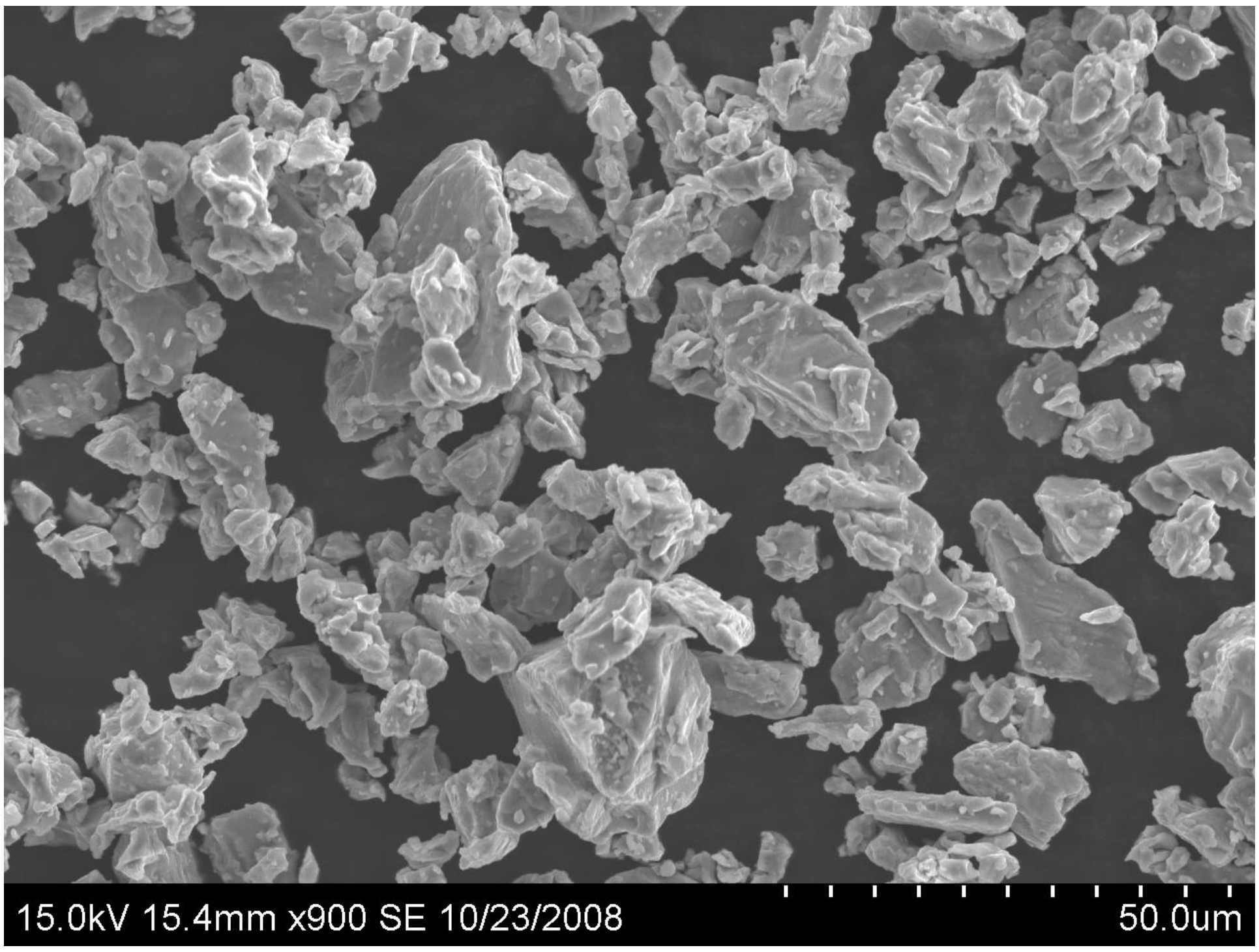

**Figure 1.** Microstructure of Ti particles

**Particle Size Distribution:** The particle size distribution was determined by laser diffraction particle size analyser (0.5-900 μm size range, 2 mW He-Ne Laser [633 nm wavelength] with 18 mm beam diameter collimated and spatially filtered to a single transverse mode [active beam length = 2.4 mm, Fourier transform lens size = 300 mm], Mastersizer S, Malvern Instruments Ltd., UK). These data were obtained for both the *complete fraction* (~1-50 μm; $d_{50}$ = ~17 μm) and a *less sedimented fraction* (~1-12 μm; $d_{50}$ = ~5 μm), as shown in Figure 2. The latter suspension was obtained by allowing the magnetically stirred complete fraction to sediment for ~5 min, followed by removal of a volume of 1.5 mL from the middle of the suspension by pipette.

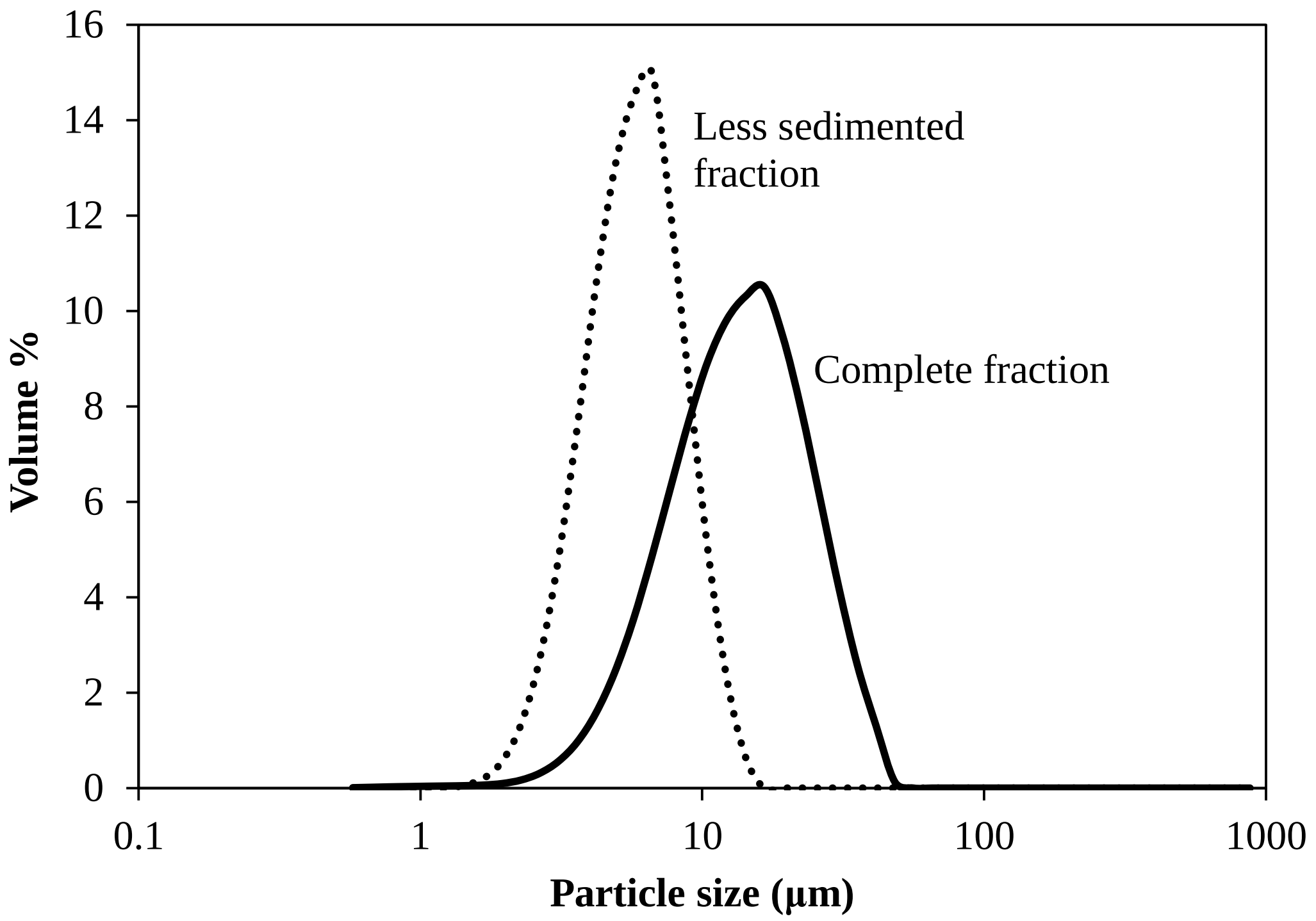

**Figure 2.** Particle size distributions for Ti particles used in the suspension for electrophoretic mobility measurements

(a) Complete fraction: As received from manufacturer, (b) less sedimented fraction: Pipetted from the complete fraction after 5 min sedimentation

**Microstructure:** The particle and deposit morphologies as well as the general appearance of the deposits were assessed by scanning electron microscopy (SEM, 15 kV accelerating voltage, secondary electron emission mode, S3400N, Hitachi High-Technologies Co., Japan).

**Electrophoretic Mobility and Electrical Conductivity:** The electrophoretic mobility and electrical conductivity were determined using a phase-analysis light-scattering zeta potential analyser (ZetaPALS; sole setting of ~10 V/cm electric field bias change with 2

Hz frequency sinusoidal wave, 0.005-30 μm size range, scattering light source [678 nm wavelength], Brookhaven Instruments Co., USA). It is likely that the thermal vibrations deriving from the use of high electric fields would be significant; thereby reducing the signal-to-noise ratio of the ZetaPALS measurements (the detection sensitivity of the ZetaPALS unit is high at low fields [19]). Consequently, the application of the commonly used low electric field of ~10 V/cm avoided this potential problem.

Test volumes of 1.5 mL each of the complete fraction (~1-50 μm) and the less sedimented fraction (~1-12 μm) were placed in a 4.5 mL standard polystyrene cuvette, agitated in an ultrasonic bath for ~1 min, and tested for electrophoretic mobility and electrical conductivity simultaneously as a function of wt% PDADMAC level (wt solid PDADMAC [in solution]/wt solid titanium). All of these background data, which are shown in Figure 3, are the averages of ten individual measurements with standard error of approximately ±0.1 μm.cm/V.s (*i.e.*, smaller than the data points).

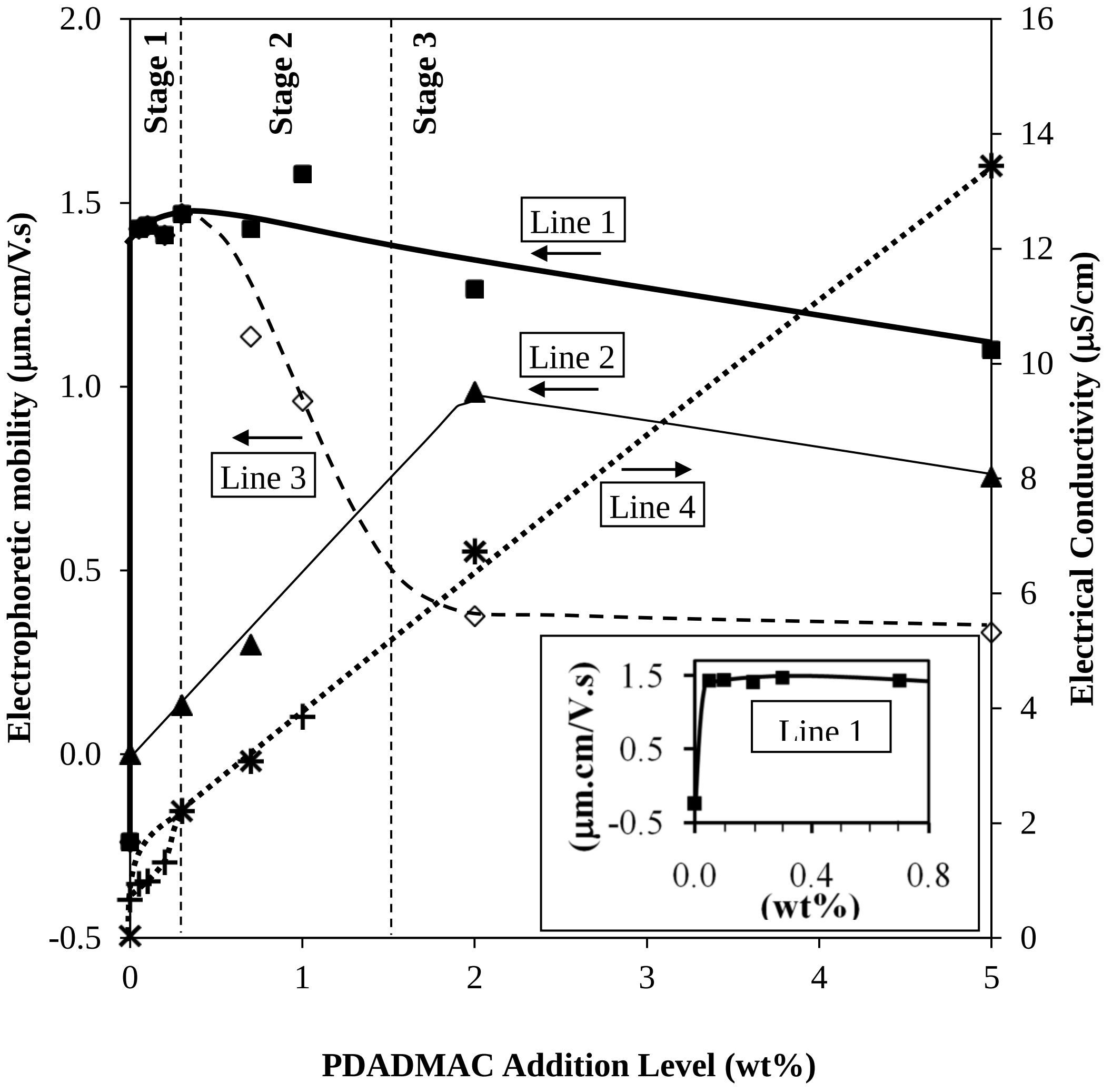

**Figure 3.** Electrophoretic mobilities ($\mu_E$) and electrical conductivities ($\sigma$) as a function of PDADMAC addition level for less sedimented fraction:

***Electrophoretic Mobility:***

| | |
|---|---|
| THICK SOLID LINE 1: | $\mu_E$ of Ti with PDADMAC additions in ethanol solution |
| THIN SOLID LINE 2: | $\mu_E$ of PDADMAC in ethanol solution |
| DASHED LINE 3: | Difference between preceding two curves |

***Electrical Conductivity:***

| | |
|---|---|
| DOTTED LINE 4: | + = $\sigma$ of Line 1 |
| | × = $\sigma$ of Line 2 |

***Inset:***

| | |
|---|---|
| CLOSE-UP: | Enlargement of Line 1, showing optimal PDADMAC level (0.3 wt%) |

The electrophoretic mobilities of the complete fraction (~1-50 μm) and the less sedimented fraction (~1-12 μm) were found to be effectively identical, indicating that the coarser fraction (~12-50 μm) sedimented rapidly, prior to measurement. Consequently, all subsequent measurements were done using the less sedimented fraction.

In the measurement of the electrophoretic mobility of solids in suspension, the resultant data generally are considered to reflect the movement solely of the particles. However, the movement of dissolved PDADMAC charging agent in the suspension and its contribution to the light scattering measurements typically are not considered despite the well established measurement of the electrophoretic mobility of polyelectrolytes in solution [20-24]. Consequently, solutions of PDADMAC in absolute ethanol were prepared and the electrophoretic mobilities and electrical conductivities were determined simultaneously as a function of wt% PDADMAC level. The analysis is based on the simplistic assumption that the amounts of light scattering from the suspended particles and dissolved polyelectrolytes are additive, thereby suggested the effect of excess PDADMAC on the electrophoretic mobility of Ti.

**EPD Set-Up:** The cathode (working electrode) or substrate consisted of SAE 1006 grade low-carbon steel (submerged dimensions 10 mm H × 5 mm W × 0.55 mm T, BlueScope Steel Ltd., Australia); the anode (counter-electrode) consisted of 304 grade stainless steel with submerged dimensions of 10 mm H × 10 mm H × 1.5 mm. The low-carbon steel substrates were hand-polished to P320 grit SiC paper (46.2 μm particle size), ultrasonically cleaned in absolute ethanol, and air-dried before deposition. All samples were used within 30 minutes of drying. The circuit consisted of mutually parallel electrodes at a fixed separation, connected by alligator clips to a d.c. programmable power supply (EC2000P, E-C Apparatus Corp., USA).

**EPD Process:** Measurements were undertaken in terms of determination of the EPD yield (weight gain/total submerged surface area) as a function of one or two variables. Since visible sedimentation was apparent immediately following mixing, each suspension was magnetically stirred for ~1 min following lowering of the electrodes into EPD suspension. After this, the voltage was applied. Each sample was removed from the suspension slowly at constant pulling rate of ~0.2 mm/s immediately after EPD ended. A low and constant pulling rate was necessary in order to minimize risk of deposit loss during removal of coated substrate from the suspension because the deposited particles were weakly bonded by electrosteric and van der Waal forces and the opposing surface tension of the liquid was of comparable scale. The weight gain was determined after EPD for each cathode by air drying for ~30 min and weighing (0.00001 g precision, BT25S, Sartorius AG, Germany).

It should be noted that these data were affected slightly by differential deposition between front and back sides of the electrode, where the deposit on the front side was greater than that on the back, at low solids loadings, deposition times, voltages, and electrode separations. There are different methods that have been used to reduce or negate this, each with different degrees of success [25,26]. In the present case, this was attempted by applying an adhesive insulating coating on the back side, although this was only partially successful. However, the deposition differentials were virtually unnoticeable after the initial stage of EPD (*viz.*, the first data point for each parameter) and, since the data for the later stages all are extrapolated to zero, these effects can be ignored.

The rationale for the selection of the experimental variables was as follows:

- **Specific Values:** The effects of PDADMAC level at the parameters of solids loading of 5 mg/mL, 5 min time point, constant voltage of 200 V and 500 V, and electrode separation of 1 cm were assessed over the PDADMAC range of 0-5 wt% (PDADMAC/Ti solids basis). A solids loading of 5 mg/mL was used as a mid-range value. A deposition time of 5 min was chosen because the division between clear supernatant and opaque sediment stabilized at and beyond this point. A bias of 200 V was selected as a minimum because an electric field of 200 V/cm was the minimum required to achieve complete areal deposit coverage on the cathode at the optimal PDADMAC addition level of ~0.3 wt%. A bias of 500 V was selected as a maximum because: (a) electrolytic corrosion of the anode commenced, which was visible in the forms of a brown colour generated in the suspension and pitting of the 304 grade stainless steel, and (b) higher voltages risked Joule heating, which could cause turbulence in the suspensions and associated deterioration of the deposit yield.
- **Solids Loading Range:** The effect of solids loading over the relatively low range of 2.5-7.5 mg/mL at the 5 min time point and electrode separation of 1 cm using three PDADMAC levels was assessed at a constant voltage of 500 V. A minimal solids loading of 2.5 mg/mL was selected in order to provide sufficient deposit mass for weighing. A maximal solids loading of 7.5 mg/mL was selected because higher values yielded samples subject to significant mass loss during removal from the remaining suspension.

- **Deposition Time Range:** The effect of deposition time over the range 1-5 min at a solids loading of 5 mg/mL, constant voltage of 500 V, and electrode separation of 1 cm using three PDADMAC levels was assessed. The minimal time point of 1 min was selected because 10 sec were required to stabilise the voltage and amperage and an additional 50 sec were required to generate sufficient deposit for weighing.
- **Voltage Range:** The effect of voltage at the solids loading of 5 mg/mL, 5 min time point, and electrode separation of 1 cm using three PDADMAC levels was assessed over the range 100-500 V, with all depositions' being done at constant voltage. The minimal voltage of 100 V was selected because it was the minimum required to produce a visible deposit yield. The maximal voltage of 500 V was selected in order to minimise anode corrosion and heating of the suspension.
- **Current Density:** The current density could not be determined accurately because the current of the EPD circuit was very low and equivalent to the resolution of the d.c. power supply, which was 1 mA.
- **Electrode Separation Range:** The effect of electrode separation over the range 0.6-2.5 cm at a solids loading of 5 mg/mL, 5 min time point, and constant voltage of 500 V using three PDADMAC levels was assessed. These separation limits were constrained by diminishing deposit yields owing to Joule heating (small separation) and decreasing electric field (large separation).

# 3. Results and Discussion

## 3.1 Apparent Effect of PDADMAC Addition on Electrophoretic Mobility of Ti Soft Particles and EPD Deposit Yield

### Conceptual Approach

The zeta potential normally is the standard parameter used to describe the surface charge of suspended particles [2]. This is calculated from the electrophoretic mobility of the particles themselves, which are assumed to be colloidal, hard, and spherical. Both zeta potential and electrophoretic mobility measurements incorporate interactive effects from the suspending medium and additives, such as excess deflocculants and charging agents. However, the DLVO theory [2] cannot explain the surface effects of electrosterically charged particles, which are known as *soft particles* [17]. The electrokinetic behaviour of polymerically charged particles is controlled predominantly by the electric potential (the Donnan potential) within the polymeric surface layer (the surface charge layer) on the underlying solid particles, as discussed in more detail subsequently. The hard particle core plus the soft saturating polymeric layer represent the soft particle. As the outer surface of the surface charge layer is approached from the inside, the initially constant (Donnan) electric potential decreases in a sigmoidal exponential fashion. This results in behaviour similar to that of the conventional double layer only in the outer diffuse layer but not in the surface charge layer. Since the DLVO theory for hard particles assumes effectively an exponential potential-distance relation [2] and the model for soft particles is different, then the concept of the zeta potential for the latter loses its physical meaning [17].

In consequence, in the present work, the electrophoretic mobility itself is reported because the zeta potential effectively assesses particulate effects only while the particle *and* the solution can be assessed separately using the electrophoretic mobility. That is, the electrophoretic mobility permits a degree of examination and decoupling of the features of particles suspended in a solution on the basis of the following arguments:

- The zeta potential assumes that the suspended particles are colloidal and spherical, neither of which is the case. The electrophoretic mobility incorporates the particle characteristics [2].
- Although the zeta potential requires knowledge of the viscosity, it normally is assumed that the pH does not alter the viscosity of the suspension or the nature of the additives, which often is not the case. The effect of viscosity is incorporated in the electrophoretic mobility [2].
- Similarly, the effect of the solids loading on the viscosity is implicit, so the preceding comments apply to the solids loading [2].
- Likewise, the effect of variable amounts of additives on the viscosity is well known, so the same considerations are applicable [2,27].
- Finally, the zeta potential applies to suspended particles only but the electrophoretic mobility allows independent assessment of the suspended particles and the dissolved species [2].

It is common in EPD and other rheological studies to attempt to optimize the surface charge of particles by varying the pH of suspensions so that the zeta potential of particles will be high and far from the isoelectric point (the pH at which the zeta potential is zero) [2]. However, owing to the corrosive natures of acids and bases used in pH adjustment and the potential for metallic corrosion, an attractive alternative method is the use of charging agents, which are not strong acids or bases.

**Data**

Figure 3 shows the electrophoretic mobilities of:

Line 1: Ti suspensions (less sedimented fraction, ~1-12 μm) in ethanol as a function of PDADMAC addition level (Ti + Total PDADMAC, including Excess PDADMAC)

Line 2 : PDADMAC solutions in ethanol as a function of PDADMAC addition level (PDADMAC)

Line 3: Difference between Line 1 and Line 2 over the PDADMAC addition level range of 0.3-5 wt% (Ti + Optimal PDADMAC)

Figure 3 also shows the electrical conductivities corresponding to Lines 1 and 2, where the only differences can be seen at PDADMAC levels ≤0.3 wt%.

The data in Figure 3 suggest the following observations and conclusions:

- The electrophoretic mobility in the absence of PDADMAC (0 wt%) was negative (see inset), which resulted from the net negative charge due to the passivating oxide layer [28]. When PDADMAC is in solution, it dissolves into a long-chain polymer terminated with a positive amine group plus free chloride. It is the positive amine group that attaches to the Ti particle, thereby reversing its surface charge from negative to positive.
- Alternatively, the role of hydroxyl groups in the ethanol and/or the aqueous PDADMAC solvent may play the dominant role in the surface charge [2,16]. In this case, the potential deprotonation of the hydroxyl group, which is attached to the passivating oxide surface, results in a net negative charge on the oxide layer.
- The optimal amount of charging agent required to assist electrophoretic mobility was quite low at only 0.05-0.3 wt% (accurate determination of the exact level using these data is not possible from these data alone; see data and inset of Figure 3.
- However, the data for the electrical conductivity of the suspensions support the preceding data through the apparent inflections at PDADMAC levels ≤0.3 wt%, which can be seen for the Ti + PDADMAC suspensions (+ data points). It would be expected that the conductivity would increase in direct proportion to the amount of free chloride in solution deriving from the PDADMAC dissociation, which is demonstrated by the data for PDADMAC solutions (× data points). That is, there is no apparent reason for there to be a connection between the saturation of the Ti surfaces with PDADMAC and the amount of free chloride in solution unless (a) the PDADMAC is not completely dissociated (unlikely) or (b) the chloride ions are localized owing to attraction to any residual free Ti surfaces (more likely).

- The optimal amount of charging agent of precisely 0.3 wt% was confirmed through measurement of deposit yield as a function of PDADMAC level, although this is discussed subsequently.
- The electrophoretic mobility of the dissolved PDADMAC would be expected in principle to be constant but it increased as the solution concentration increased, reaching a maximum at ~2 wt%, slightly decreasing thereafter. These effects probably result from the influence of two competing mechanisms, both of which increase as the PDADMAC concentration increases: (a) low PDADMAC levels – increasing proximity of the molecules, consequently increasing the alignment owing to mutual repulsion and the effect of the directional electric field, and resultant greater streamline flow and (b) high PDADMAC levels – increasing viscosity. The inflection corresponds to the point at which the second mechanism begins to dominate over the first.
- It may be noted that, following saturation, Line 1 (Ti + Excess PDADMAC) decreased linearly while the PDADMAC curve altered significantly. In this case, the former data are likely to result from a variation of the two competing mechanisms: (a) low PDADMAC levels (<2 wt%) – dominance of the scattering effect of the large opaque Ti particles compared to the small transparent PDADMAC molecules (*viz.*, large differences in measured electrophoretic mobilities) and (b) high PDADMAC levels (≥2 wt%) – dominance of the viscosity and greater drag on the Ti particles (*viz.*, small differences in measured electrophoretic mobilities).
- Since the optimal amount of PDADMAC for Ti saturation apparently is low at 0.3 wt%, then most of the suspensions had free PDADMAC. As mentioned above, the effect of the light scattering by the excess PDADMAC on the electrophoretic mobilities (at ~0.3-5 wt%) has been subtracted in order to assess the electrophoretic mobilities of the optimally charged Ti particles in the absence of the extraneous effects of the excess PDADMAC, as shown by Line 3 in Figure 3. These data can be described in terms of three ranges (Stages 1-3), which have been confirmed by direct experimental measurement, as clarified subsequently.

- Since there was no difference between the data for the less sedimented and complete fractions, it is clear that the larger particles sedimented vertically while the finer particles moved horizontally under the effect of the relatively low electric field of ~10 V/cm, which was oriented for horizontal mobility.

Figure 4 shows the deposit yield of Ti particles over the submerged surface area of the cathode at 200 V/cm and 500 V/cm as a function of PDADMAC level. These data differ only in the scale of the deposit yields, where, as expected, the higher electric field resulted in greater deposit yield owing to the greater driving force. The consistency of the inflections within Figure 4 and in comparison to the same inflections in Figure 3, which shows three sets of data, two of which are independent, is significant. That is, there are three regions that can be differentiated: (a) a rapid increase in the deposit yield up to a maximum at the optimum of 0.3 wt% PDADMAC (Stage 1), (b) a relatively rapid decline in deposit yield up to ~1.5 wt% PDADMAC (Stage 2), and (c) a gradual decline in the deposit yield to 5 wt% PDADMAC (Stage 3). These data suggest the following observations and conclusions:

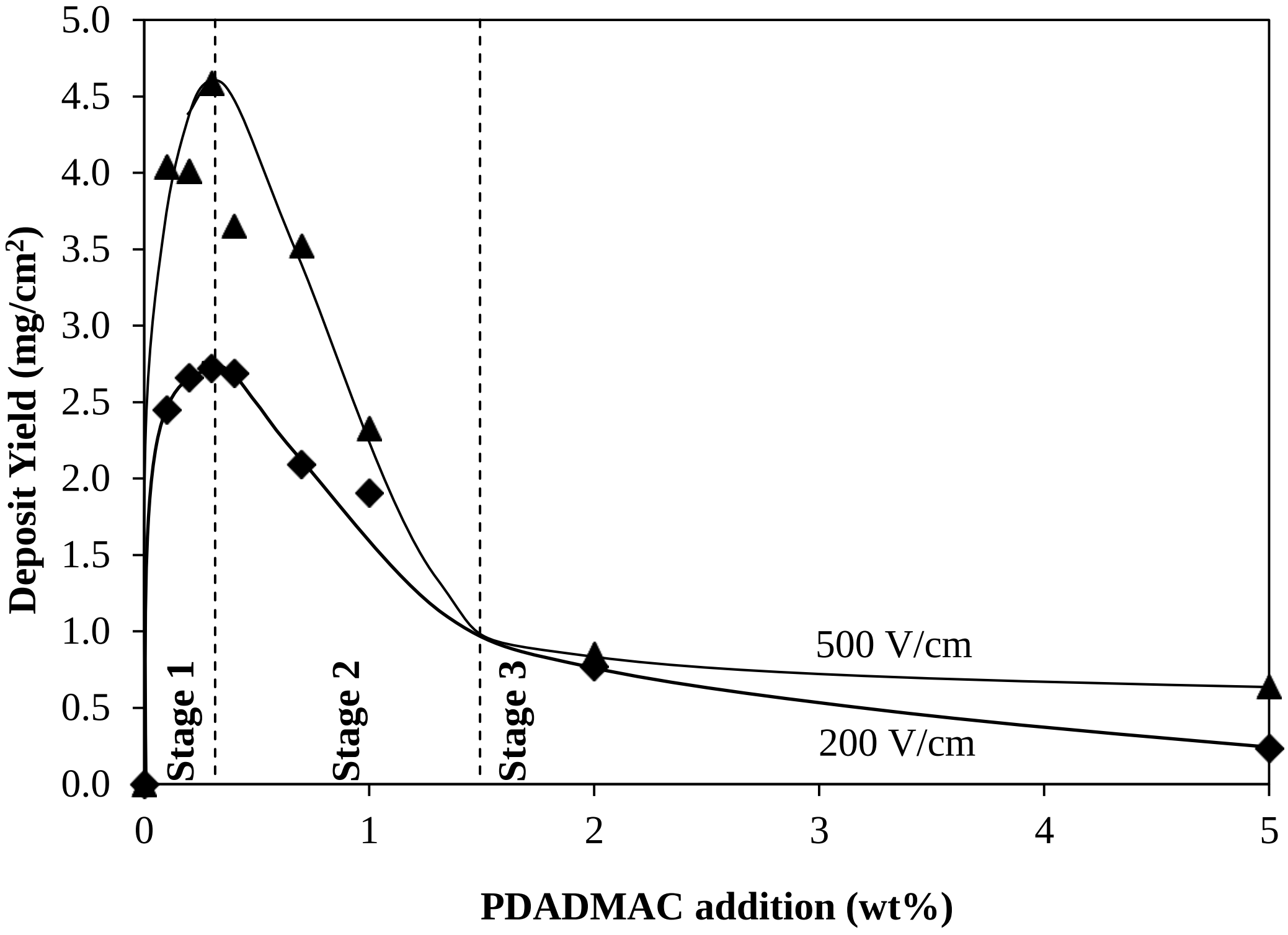

**Figure 4.** Dependence of the deposit yield on the PDADMAC addition level for complete fraction (solids loading = 5 mg/mL, deposition time = 5 min, applied electric field = 200 and 500 V/cm)

- **Stage 1:** The relatively rapid increase in deposit yield reflects the data of Figure 3 in that saturation of the particle surfaces by the charging agent was reached at a relatively low level of PDADMAC. The ambiguity in the precise level of optimal PDADMAC addition in Figure 3 is removed by the data in Figure 4 because they clarify the maximal deposit yield, especially at 500 V/cm, as being at 0.3 wt%. This maximum represents a threshold or balance between the Ti surface area and the amount of PDADMAC necessary to saturate it, which is independent of the electric field (Figure 4) and dependent entirely on the physical relationship between the available particle surface area and the volume and packing of the saturating polyelectrolyte. Hence, the inflections for 200 V/cm and 500 V/cm in Figure 4 occur at the sole optimal amount of charging agent. This observation is useful because it demonstrates that measurement of the electrophoretic mobility of the Ti in low field, as shown in Figure 3, is applicable to electrophoretic deposition at high field, as shown in Figure 4. Since Figure 2 shows that the particle size of the less sedimented fraction was ≤12 μm, this provides the useful observation that low electric fields for the measurement of the electrophoretic mobility can be applicable to soft particles as large as 12 μm.

- **Stage 2:** The relatively rapid decline in deposit yield resulted from the progressively decreasing differential between the electrokinetics of the optimally charged Ti particles (higher mobility) and the PDADMAC (lower mobility), both of which carry a net positive charge. This can be explained by two divergent scenarios:

  (a) **Mobility Effect: Compression of Diffuse Layer:** With the increasing ionic (positive and negative) concentration from excess PDADMAC, compression of the diffuse layer [2] surrounding the optimally charged Ti soft particles during Stage 2 (Figure 5) reduces the electrophoretic mobility (Line 3 in Figure 3). This would reduce the deposition rate and thus decrease the deposit yield, as confirmed in Figure 4. It should be noted that the only significant differences between the hard particles of DLVO theory and the soft particles of the present work are: (i) the former considers only electrostatic and van der Waals forces for colloidal particles [2] while the latter accommodates electrosteric forces in non-colloidal particles as well [17] and (ii) the former includes three distinctly different electric potentials (surface,

Stern, and zeta) while the latter includes the approximately equal surface and Donnan potentials.

(b) **Electrode Effect: Interposing Polymer-Rich Layer Growth:** With increasing amount of PDADMAC, a greater proportion of polymer-rich material would interpose the cathode surface and Ti particles. This would serve to reduce the adhesive strength between the steel cathode and Ti particles since the negatively charged oxide layers on both metals can be assumed to be better bridged by a single positive amine molecule as compared to a thick polymer-rich layer. It also would serve to deposit an insulating layer on the electrode, thereby reducing the rate of deposition. Hence, the effect of excess polymer would be to reduce both deposition rate and effectiveness of adhesion, the latter of which would enhance dislodgement of the Ti particles from the cathode surface during the EPD process. Also, the competition between the electrophoretic mobilities of PDADMAC and the Ti particles (as shown in Figure 3) can be seen to alter in favour of PDADMAC as its level increases.

- **Stage 3:** The gradual decline in deposit yield at higher PDADMAC levels and the associated inflection between the two rates of deposition suggest two complementary scenarios:

  (a) **Mobility Effect: Donnan Potential Constancy:** In the limit of high ion concentration (when PDADMAC level reaches the inflection at 1.5 wt%), the surrounding diffuse layer of soft particles and its corresponding electrical potential were diminishing to zero, as shown in Figure 5. As the Donnan potential within the surface charge layer of soft particles is only slightly affected by the increasing ion concentration [17] during Stage 3, the electrophoretic mobility of the soft particles showed near-zero decrease. Therefore, the electrophoretic mobility and deposit yield should approach a constant level, as is shown in Figures 3 and 4.

(b) **Electrode Effect: Interposing Polymer-Rich Layer Growth:** At the inflection at ~1.5 wt% PDADMAC, a threshold thickness of interposing polymer-rich layer is achieved (as discussed subsequently). This inflection is associated with the establishment of one or more of the following: (i) critical thickness for adhesion (Stage 2), (ii) critical electrical resistance to cathode-Ti attraction (Stage 2), and (iii) establishment of dominance of the electrophoretic mobility of PDADMAC over that of Ti particles (Figure 3). All three of these phenomena are suggestive of the cause of the inflection between Stages 2 and 3 and they support the conclusion of a change in deposition mechanism at the threshold of ~1.5 wt% PDADMAC.

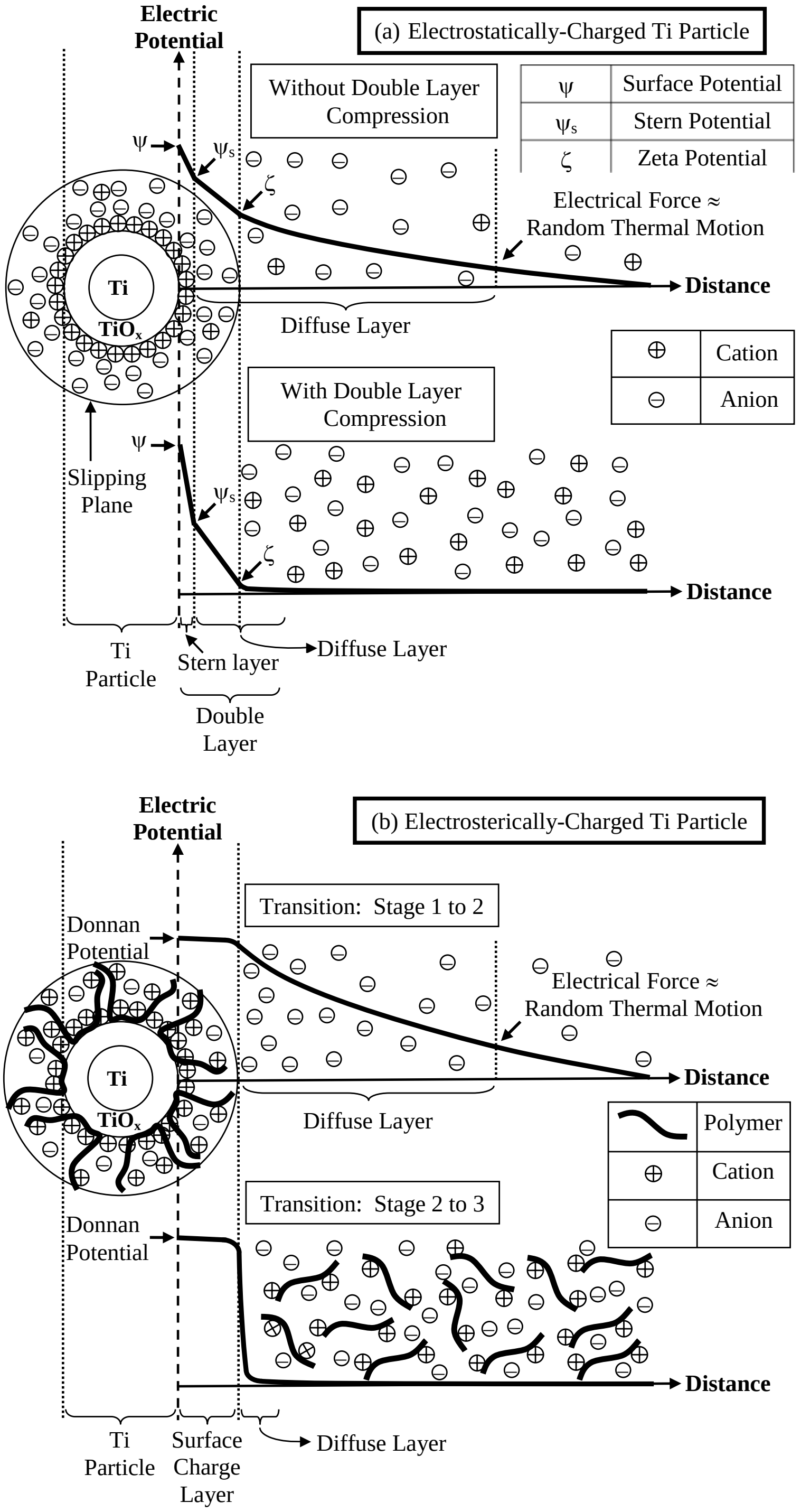

**Figure 5.** Schematic illustration showing comparison between: (a) electrostatically-charged particle (based on conventional DLVO theory for electrolytes [29]), and (b) electrosterically-charged particle (based on present work for polyelectrolytes; transitions for Stages 1 to 2 and 2 to 3 refer to Figures 3 and 4, respectively)

The correspondence of the inflections and trends of the four sets of data, three of which are independent, in Figures 3 and 4 tend to self-support the preceding conclusions. These correspondences are not surprising because: (a) Line 3 in Figure 3 is for optimally charged Ti particles, which decouples the effect of the excess PDADMAC and (b) since the weight of the deposited polymer is significantly less than that of the deposited Ti particles, then Figure 4 also decouples the effect of excess PDADMAC.

Other relevant issues concerning the data in Figures 3 and 4 are as follows:

- **Dislodgement of Deposit:** Figure 6 shows that EPD (200 V/cm) during Stages 1 and 2 (0.3 wt% and 0.7 wt% PDADMAC, respectively) was characterized by some dislodgement of the deposits. While the microstructures for PDADMAC levels of 0.1, 0.2, 0.3, and 0.4 wt% showed only minor losses, those for PDADMAC levels of 0.7 wt% showed more substantial dislodgement. The latter resulted from the gradual reduction in adhesion with increasingly excess PDADMAC, as discussed for Stage 2.

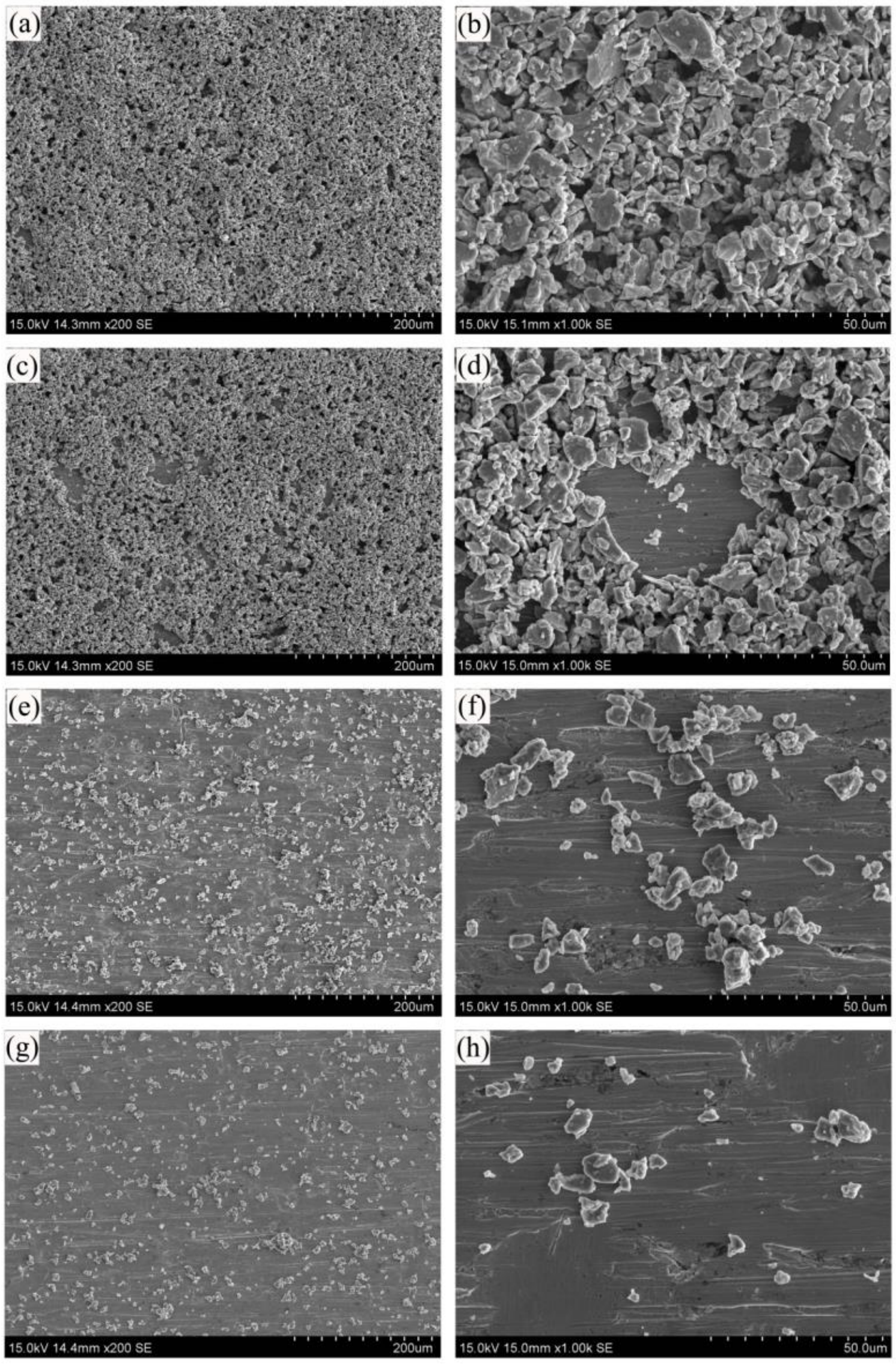

**Figure 6.** SEM micrographs of Ti deposits using suspensions with four PDADMAC addition levels for complete fraction

(a), (b):  0.3 wt%; (c), (d):  0.7 wt%; (e), (f):  2 wt%; (g), (h):  5 wt%

(a), (c), (e), (g):  x200 magnification; (b), (d), (f), (h):  x1000 magnification

(solids loading = 5 mg/mL, deposition time = 5 min, applied voltage = 200 V, electrode separation = 1 cm)

- **Microstructural Change:** Figure 6 shows that the microstructures associated with Stages 2 and 3 (0.7 and 2 wt% PDADMAC, respectively) had significantly different areal densities of deposition, which can be attributed to the decline in electrophoretic mobility indicated by Line 3 of Figure 3. The use of Line 1 does not allow this distinction to be made as the change in mobilities for Ti + Excess PDADMAC is small. Figure 6 also can be interpreted in terms of: (a) the establishment of the previously discussed polymer-rich layer of critical thickness and/or (b) electrokinetics dominated by PDADMAC rather than Ti particles (where Figure 3 shows that the electrophoretic mobility of PDADMAC exceeds than that of Ti + optimal PDADMAC during Stage 3).

- **Viscosity Effect at Lower Excess PDADMAC Level:** The effect of increasing viscosity from increasing PDADMAC level is not considered to be responsible for this threshold (from Stage 1 to Stage 2) because, first, the higher viscosity and associated drag would be expected to retard the deposition of the smaller particles, which is not the case (Figure 6), and, second, a gradually increasing viscosity would not be expected to result in an inflection in the data, whereas a change in mechanism would.
- **Viscosity Effect at Higher Excess PDADMAC Level:** However, in relation to Figure 3, it would be expected that the viscosity should increase with increasing PDADMAC level. Hence, within Stage 3, the slight and gradual decrease in the electrophoretic mobility can be attributed to the effect of the viscosity and the consequent drag on the small Ti particles, which are capable of being moved by the low electrical field of ~10 V/cm. However, when the data are decoupled and the electrophoretic mobilities of the Ti + Optimal PDADMAC are examined (Line 3), it is clear that there is no effect from the viscosity.

- **Depositable Particle Size:** Electrophoretic deposition of the complete fraction at the high electric fields of 200 and 500 V/cm for 5 min resulted in: (a) rapid sedimentation of the coarse particles (~20-50 μm), (b) initial deposition of a mixture of large (~12-20 μm) and small particles (~1-12 μm), and (c) subsequent deposition of small particles (~1-12 μm). This gradual time-dependent deposition is discussed in Section 3.2.
- **Contamination from Anode Corrosion:** Another potential factor in the overall decrease in deposit yield with increasing PDADMAC level resulted from contamination owing to the progressive corrosion of the stainless steel anode. This would have affected the pH, electrical conductivity, and/or viscosity. However, this is unlikely to be the case since the data trend in Figure 4 is not consistent with any of these mechanisms. For example, if corrosion were responsible, then it would be expected that increasing amounts of chloride ion from the increasing PDADMAC levels would generate an increasing rate function rather than the observed decreasing rate function.

## 3.2 Effect of Solids Loading, Deposition Time, Applied Voltage, and Electrode Separation on Deposit Yield

The deposit yield can be considered in light of the well known relation proposed by Hamaker [30]:

$$W = f \int \left[ \mu \left( \frac{V}{d} \right) AC \right] dt$$

Where: W = Weight of deposit yield (g)

f = Efficiency factor (f ≤ 1; f = 1 if all particles are deposited) (unitless)

μ = Electrophoretic mobility of particles (μm.cm/V.s)

V = Applied voltage (V)

d = Distance between electrodes (μm)

A = Surface area of the substrate used ($cm^2$)

C = Solids loading ($g/cm^3$)

t = Deposition time (s)

Previous work [26] on the EPD of alumina colloidal particles in isopropanol deposited on stainless steel substrates indicated that the most effective means of increasing the deposit yield are, in order of effectiveness, increasing the: (a) solids loading, (b) deposition time, (c) applied voltage, and (d) distance between the electrodes (electrode separation). However, these observations were made for colloidal ceramic materials, which generally are relatively easy to deposit. Metals are much more difficult to deposit owing to their lower electrophoretic mobilities, which derive from their lower surface charges and larger particle sizes. Figures 3 and 4 indicate that modification of the electrophoretic mobility through the use of a charging agent can overcome these obstacles. In this sense, increasing the electrophoretic mobility must be considered the primary factor in increasing the deposit yield of metals.

The concurrent effects of the electrophoretic mobility of the complete fraction of the suspensions (~1-50 μm) and the four other variables (a-d) are shown in Figure 7 (solids loading with the corresponding electrical conductivity), Figure 8 (deposition time), Figure 9 (applied voltage), and Figure 10 (electrode separation with the corresponding electric field). These data are for three near-optimal levels of charging agent, as follows:

Undersaturation coverage – 0.2 wt% PDADMAC

Saturation coverage – 0.3 wt% PDADMAC

Oversaturation coverage – 0.4 wt% PDADMAC

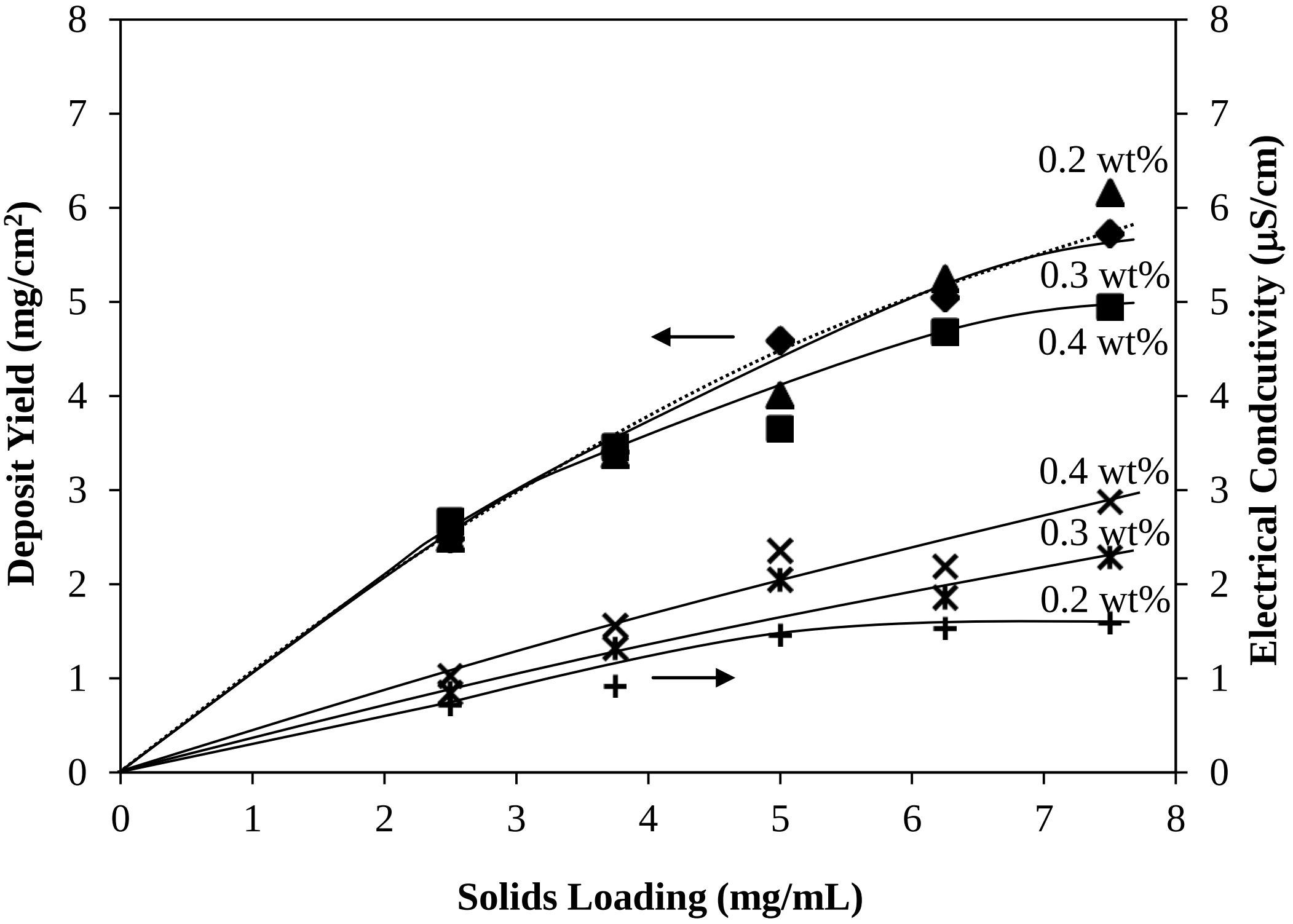

**Figure 7.** Deposit yield as a function of solids loading for three PDADMAC addition levels with their corresponding electrical conductivities for complete fraction (deposition time = 5 min, applied voltage = 500 V, electrode separation = 1 cm)

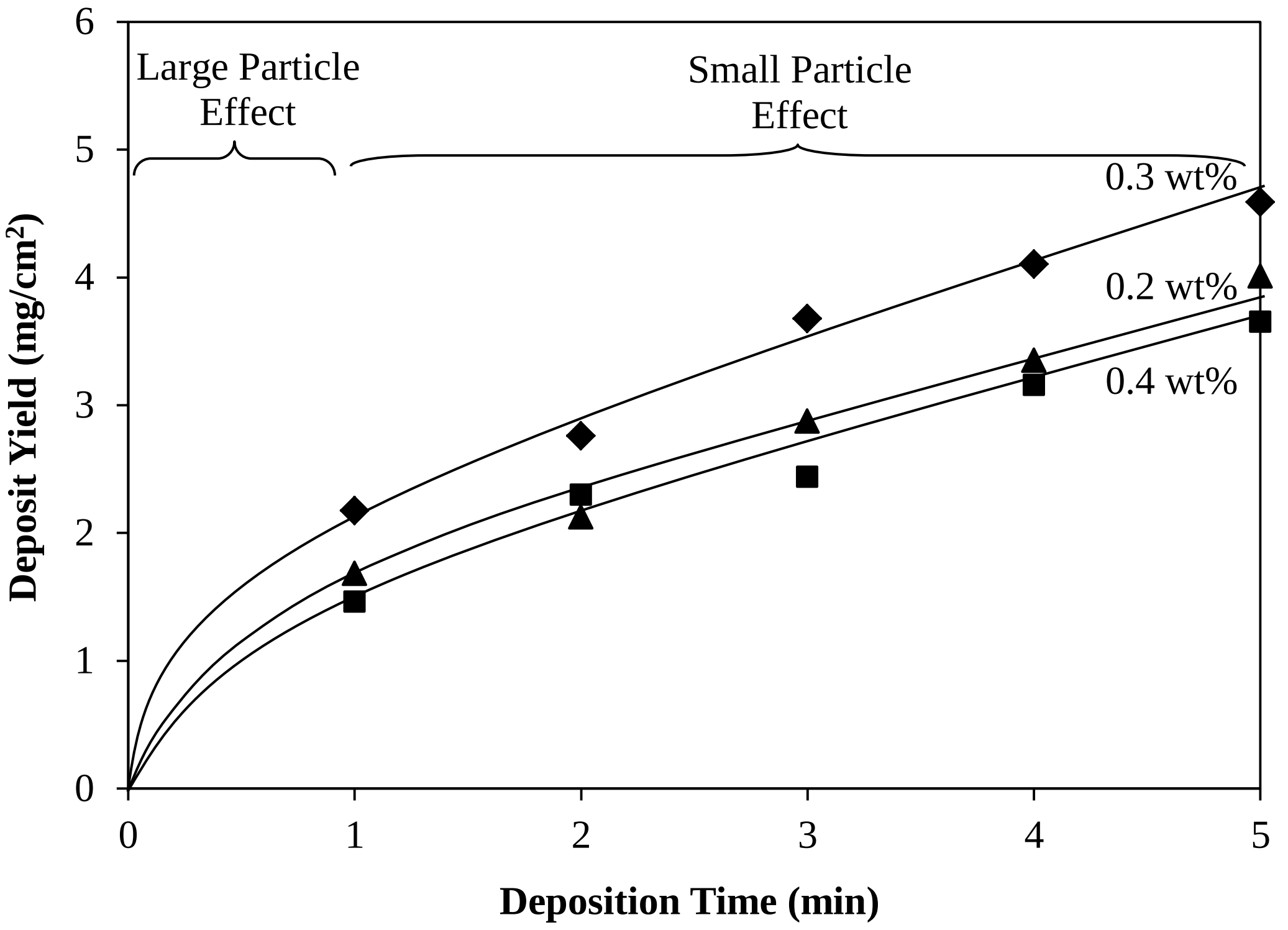

**Figure 8.** Deposit yield as a function of deposition time for three PDADMAC addition levels for complete fraction (solids loading = 5 mg/mL, applied voltage = 500 V, electrode separation = 1 cm)

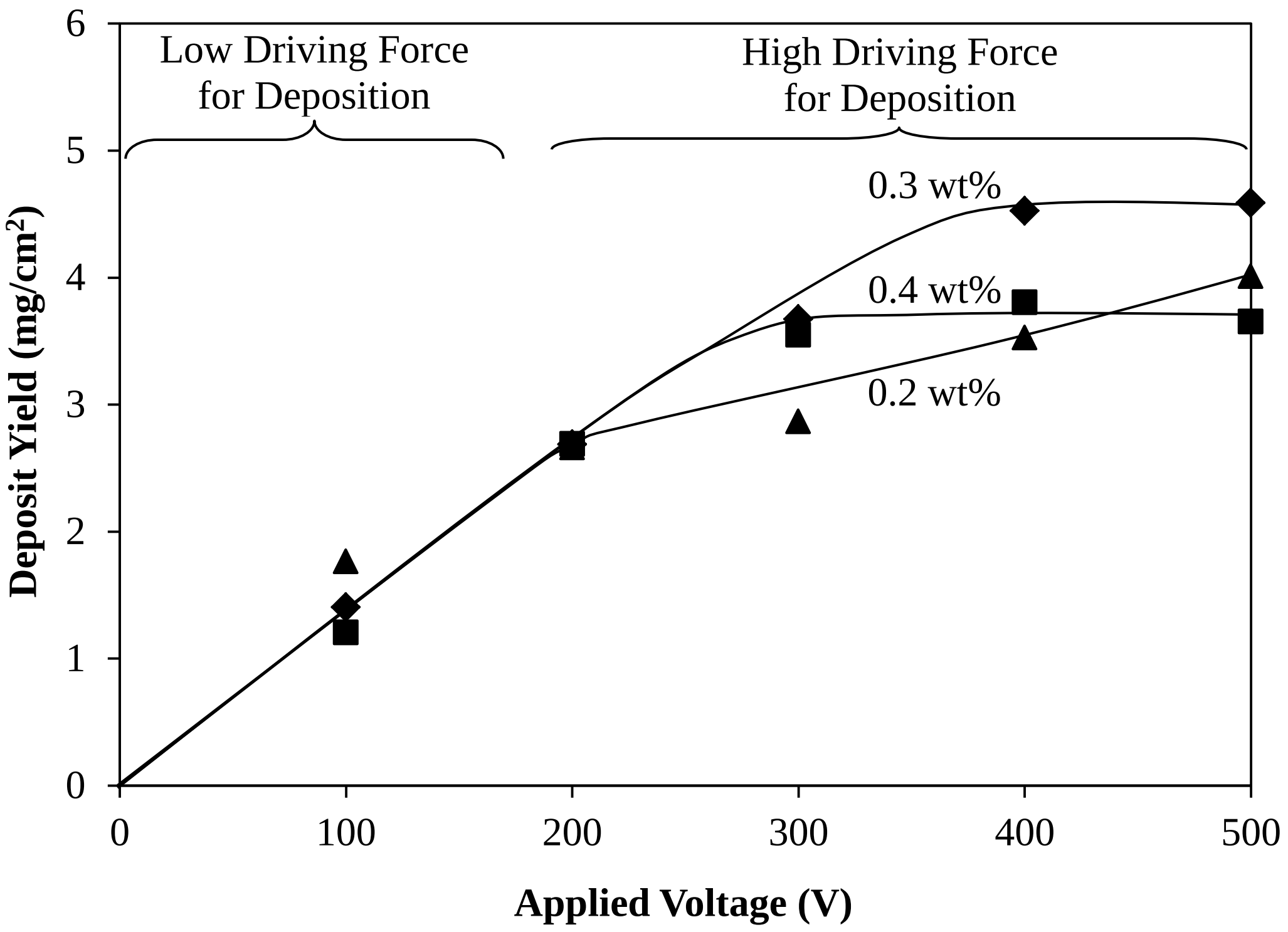

**Figure 9.** Deposit yield as a function of applied voltage for three PDADMAC addition levels for complete fraction (solids loading = 5 mg/mL, deposition time = 5 min, electrode separation = 1 cm)

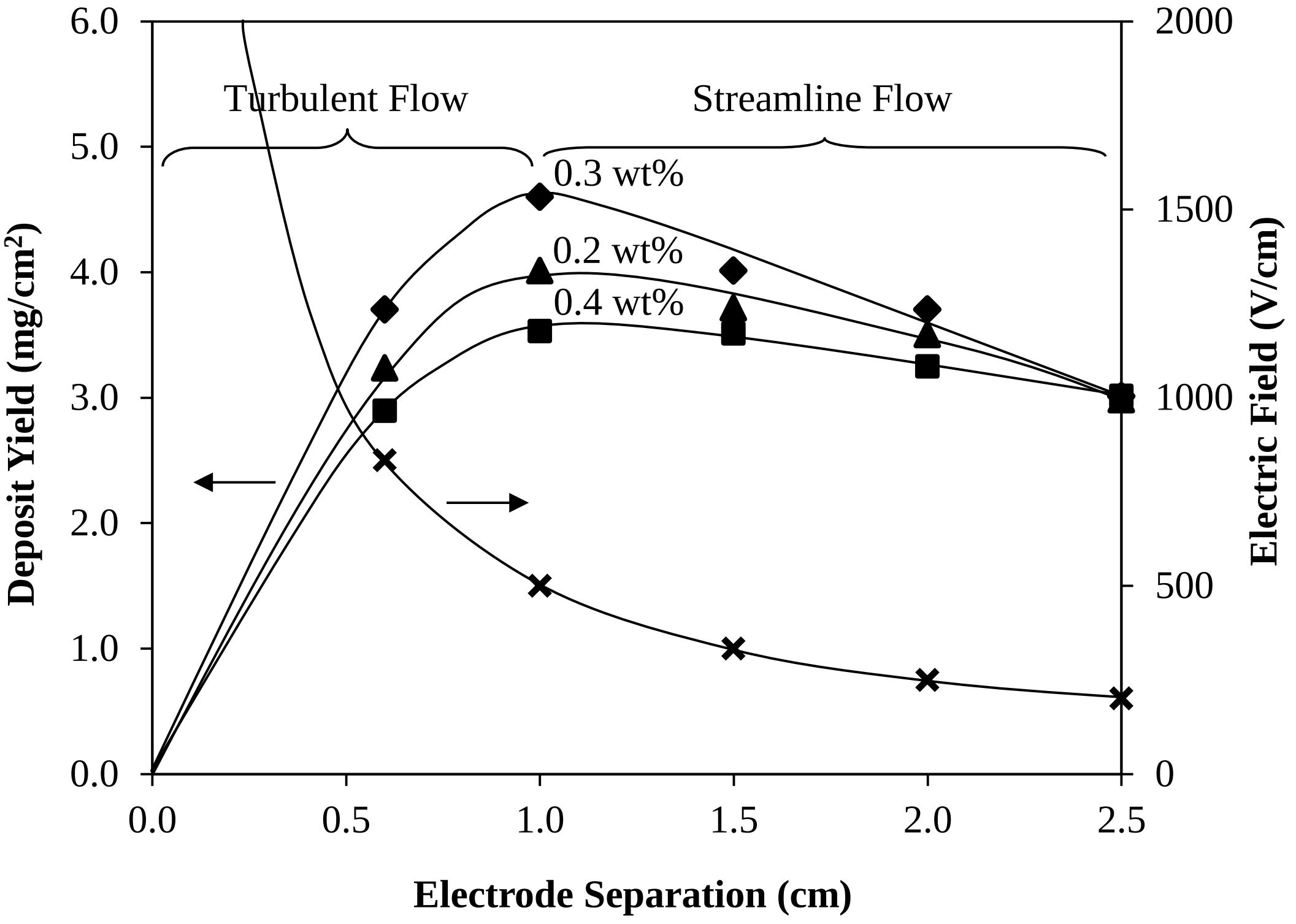

**Figure 10.** Deposit yield as a function of electrode separation for three PDADMAC addition levels for complete fraction; the electric field corresponding to the electrode separation also is shown (solids loading = 5 mg/mL, deposition time = 5 min, applied voltage = 500 V)

The data in Figure 7 (solids loading with the corresponding electrical conductivity) suggest the following observations and conclusions:

- Figure 3 shows that, for a constant solids loading, the electrical conductivity increases with increasing PDADMAC levels. These data confirm this over the range of solids loading investigated.
- Figure 4 shows that the deposit yields for the three near-optimal levels of charging agent are essentially indistinguishable. These data confirm this over the range of solids loading investigated.

- The most obvious effect is that the deposit yield increases logarithmically as a function of solids loading. Hamaker [30] explained this in terms of the pressure applied on initially deposited particles by those deposited later, compressing the particles so that van der Waals and steric forces can overcome the electrostatic repulsion forces between the particles. However, it should be noted that a simultaneous effect is that the electrical conductivity of the electrode + deposit progressively decreased with increasing deposit thickness owing to the point contacts between the Ti particles and the insulating layer of the polymeric component of the PDADMAC on the deposited particles; these effectively reduced the electric field at the outer deposit layer, thereby reducing the driving force for further deposition.
- A common interpretation of reduced ceramic EPD deposit yields at higher applied voltages is in terms of an increasing electrical resistance barrier due to the deposited insulating particles on the electrode, which reduces the electric field strength and thus the driving force for further deposition [31]. In the present case, it might be assumed that the deposit of conductive metallic particles should not decrease significantly the electrical resistance. It may be noted that the overall resistivity of the substrate + deposit system increased with increasing deposit yield owing to: (a) the higher electrical resistivity of titanium ($3.9\times10^{-7}$ Ωm [32]) compared to that of low-carbon steel ($1.3\times10^{-7}$ Ωm [33]), (b) the resistance from the passivating oxide on the Ti particles, and (c) the aforementioned resistance from the PDADMAC deposited on the substrate.
- Owing to these factors, a packing gradient is likely to exist through the thickness of the deposit, with the outer layers' being the most weakly bonded. As the layer thickness increased, the packing density and associated cohesiveness decreased, making the outer layers subject to easier loss during removal from the remaining suspension and subsequent handling. This potential was demonstrated when solids loadings >7.5 mg/mL (at the data cut-off in Figure 7) were used. If the effects of pressure and electric field dominated, then the data should have levelled off to a constant value; if the effects of mechanical loss dominated, then the data should have shown a maximum (at 7.5 mg/mL). Since the latter was the case (the data to 25 mg/mL are not shown), the smoothness of the data in Figure 7 suggest the following:

≤7.5 mg/mL: Pressure and electric field effects dominate

>7.5 mg/mL: Mechanical loss effect dominates

The data in Figure 8 (deposition time) suggest the following observations and conclusions:

- The present trend of approximately logarithmic data is consistent with those for electrically insulating ceramics [31] and conducting metals [11,12]. This trend, which corresponds to the condition of constant voltage, which was used, results from (a) the decreasing electric field as the electrical resistance of the substrate + deposit increases and (b) the decreasing level of solids loading in the suspension as deposition proceeds.
- Examination of Figures 8 and 11 shows the deposition of the large and heavy particles (up to ~20 μm). These particles were in close proximity to the electrode, with little horizontal distance to travel for deposition during the early stage (first 1 min) and little time to sediment vertically. While smaller particles (~1-12 μm) were depositing continuously during all stages, the larger particles (~12-50 μm) further from the substrate were sedimenting vertically, which resulted in the effective segregation of the particles such that, at the later stages (after 1 min), only smaller particles (~1-12 μm, typically ~5 μm in size) were available to deposit.

- The more linear later stage (>1 min) of the data results from dominance of the deposition of a relatively large supply (*i.e.*, constant concentration) of small particles relative to the amount of deposit (the deposit yield represents ~4 wt% of the total amount of solids). The linearity of the data suggests that, despite the particle size range (~1-12 μm) and median size (~5 μm) of these particles (Figures 6 and 11), they deposited similarly to colloids.
- Since Figure 7 shows a parallel behaviour for the deposit yield and electrical conductivity as a function of solids loading and Figure 3 shows a direct relation between the electrical conductivity and PDADMAC level, then these two figures indicate that the deposit yield should increase with increasing PDADMAC level. However, examination of the three close PDADMAC levels in Figure 8 does not support this.
- Although Figure 7 (plotted in terms of the solids loading at a constant deposition time) indicates that increasing electrical conductivity results in increasing deposit yield, Figure 8 (plotted in terms of the deposition time at a constant solids loading) does not support this conclusion. This is explained in Figure 4, where the deposit yield depends critically on the PDADMAC level. Hence, 0.3 wt% PDADMAC also shows the highest deposit yields, which result from the maximal level of the charging agent and resultant optimal saturation of the Ti particle surfaces. So the optimal amount of PDADMAC would depend on the solids loading. That is, although the PDADMAC levels of 0.2 and 0.4 wt% do not appear to be differentiable in Figure 8, they are in Figure 4.

- In Figure 8, the importance of the PDADMAC level was established effectively immediately (≤1 min). During this initial period, both large and small particles were deposited owing to their proximity to the cathode, where the majority of the weight gain derived from the large particles (large particle effect). At the later deposition times (>1 min), the weight gain from the deposition of small particles was predominant (small particle effect). These observations are demonstrated clearly in the SEM images in Figures 11. The data for 0.3 wt% PDADMAC show the greatest deposit yield because the coverage of the Ti particles was optimally saturated, thereby providing maximal adhesion through the effect of the polymer. The lower deposit yields for 0.2 wt% PDADMAC resulted from incomplete coverage and therefore less adhesion. The lower deposit yields for 0.4 wt% PDADMAC resulted from the presence of excess PDADMAC in solution and therefore weaker adhesion from the additional interposing polymer, as explained in the description of Stage 2.
- The data in Figure 8 also reflect the influence of the electrophoretic mobilities of the particles. Examination of Figure 3 demonstrates the importance of discriminating between the data for the suspension (Line 1) and those for the optimally saturated Ti particles (Line 3). That is, examination of Line 1 alone, which would be the common practice, does not allow a clear conclusion concerning an optimal amount of charging agent; additional data in the form of those given in Figure 4 are required. The amount of effort to generate the data in Line 2 for the PDADMAC solutions is considerably less than that required to generate those in Figure 4.

- It is not surprising that Figures 3, 4, and 8 are consistent in terms of the effects of the electrophoretic mobility on the deposit yield, where 0.3 wt% PDADMAC is optimal but those for 0.2 wt% and 0.4 wt% PDADMAC are inferior but not differentiable.

The data in Figure 9 (applied voltage) suggest the following conclusions:

- There is a limited amount of published data on the deposit yield as a function of voltage and these are contradictory. Ceramic deposits have been observed to show linear trends [10,34] and metallic deposits (with charging agents) have shown exponential trends [10,12]. The data for the metals also showed maxima, which were attributed to electric arcing [12] and loss of agglomerated volumes from the deposit surface [10]. Other reports have suggested that the maxima result from turbulence at high voltages [31] and unstable voltage and current density [35]. The present data are similar but not sufficiently distinctive to allow direct comparison with published data.
- The present data can be divided into regions of low and high driving forces for deposition. At the two lowest voltages, the interaction between the electric field and the surfaces of the particles charged with PDADMAC is so low that the results cannot be differentiated.
- At the higher voltages, the higher electric field is sufficiently strong to interact with the surfaces of the particles, so the data diverge.
- Comparison within Figure 11 (c)-(d) for 500 V and (e)-(f) for 100 V complements these comments. That is, the lower voltage yielded incomplete coverage while the higher voltage yielded complete coverage.

- It can be seen that the inflections in these data occur at different voltages:

| | |
|---|---|
| 0.3 wt% PDADMAC | Inflection commences at 400 V and the data level out |
| 0.4 wt% PDADMAC | Inflection commences at 300 V and the data level out |
| 0.2 wt% PDADMAC | Inflection commences at 200 V but the data do not level out |

These data again can be explained in terms of the degree of coverage of the particle surfaces by the PDADMAC. With 0.3 and 0.4 wt% PDADMAC, surface coverage was complete, although the former was optimal and so exhibited the highest deposit yield and inflection at the highest voltage. The apparent absence of maxima allows the speculation that the curves become approximately constant owing to the formation of a threshold insulating thickness beyond which the electric field generated by the applied voltage has little or no effect. Hence, at 0.4 wt% PDADMAC, the greater amount of polymer interposed the Ti particles and the cathode allowed establishment of this insulating layer at a lower voltage. Owing to the possibilities of thermal and current instabilities and agglomerate loss at higher voltages [10,12,31,35], the eventual observation of maxima (in this case, flat) is inevitable. With 0.2 wt% PDADMAC, the surface coverage was incomplete, which allows closer packing of the particles and thus effectively thinner coatings. Hence, it takes longer to reach the ultimate threshold thickness of the insulating layer. The other two PDADMAC levels had saturated Ti particles and, volume-wise, these were indistinguishable.

- Examination of Figure 7 shows that, if a higher solids loading is used, a higher deposit yield can be obtained using the same conditions of time, voltage, and electrode separation. However, Figure 9 appears to contradict this by indicating that, with increasing voltage at a constant solids loading of 5 mg/mL, a maximal achievable deposit yield is established. This apparent conflict can be resolved by examination of the corresponding data in Figures 7 and 9 (solids loading of 5 mg/mL and 500 V), which show that the deposit yields are consistent. The data in Figure 7 can be interpreted in terms of the solids loading. That is, at higher solid loadings, EPD occurs more quickly and the pressure exerted on the deposit is greater, which still could result in the same threshold deposit thickness, just obtained in a shorter time (note the non-linear time dependence shown in Figure 8).

- Further, it has been suggested that the flat maxima in Figure 9, which may correspond to a transition from streamline to turbulent flow of the dispersed Ti particles, can be suppressed (for a system consisting of a dispersion of rigid polymeric particles) using higher solids loadings [36].

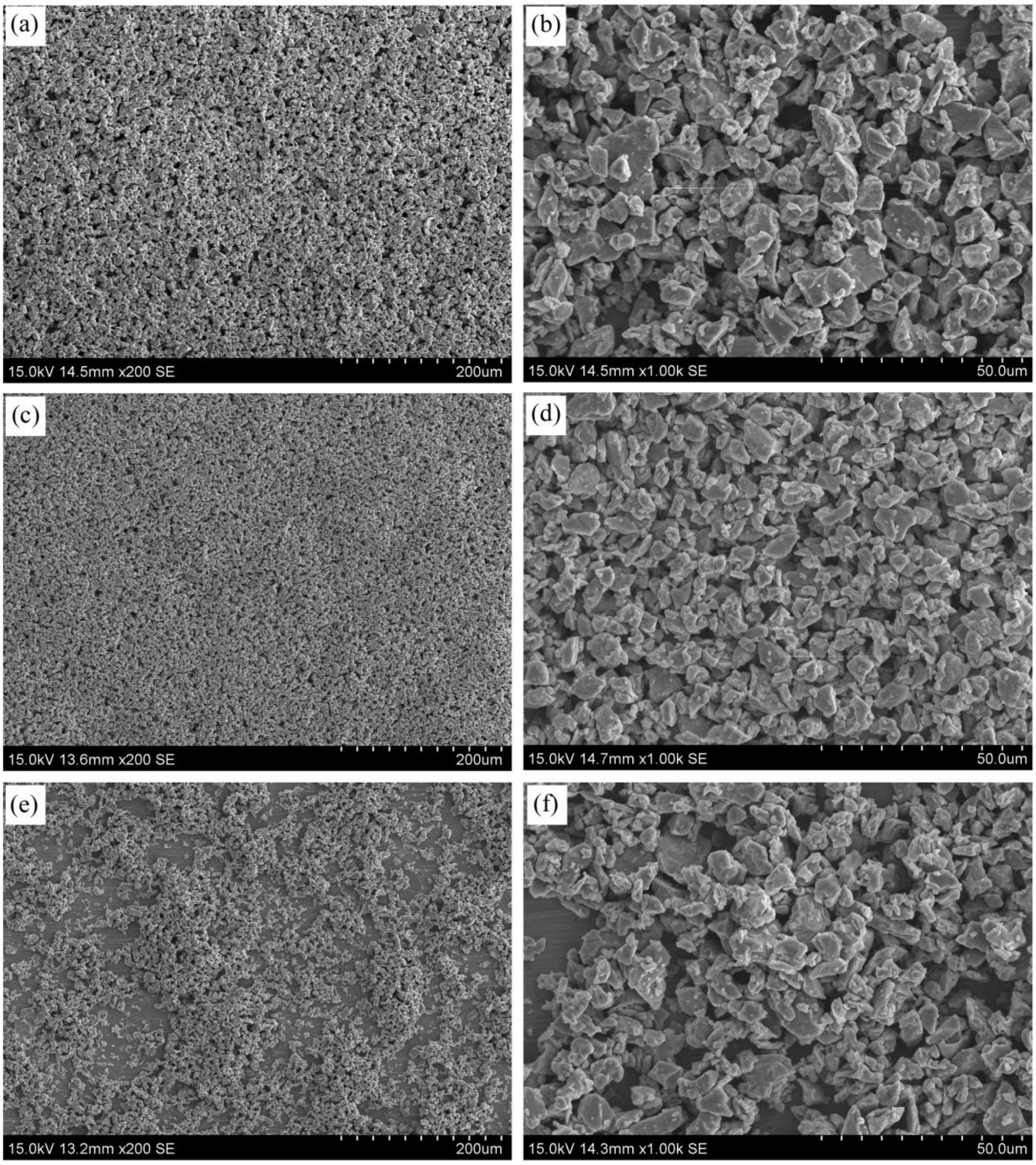

**Figure 11.** SEM micrographs of Ti deposits at various applied voltages and deposition times for complete fraction (PDADMAC addition level = 0.3 wt%, solids loading = 5 mg/mL, electrode separation = 1 cm)

(a), (b): 500 V and 1 min; (c), (d): 500 V and 5 min; (e), (f): 100 V and 5 min

(a), (c), (e): x200 magnification; (b), (d), (f): x1000 magnification

The data in Figure 10 (electrode separation with the corresponding electric field) suggest the following observations and conclusions:

- At low separation (<1.0 cm), the electric field is very high, so Joule heating and associated suspension turbulence occur. Thus, increasing the separation over this range decreases this effect, causing the deposit yield to increase. This region is characterized by turbulent particulate flow.
- At higher separations (≥1.0 cm), more typical EPD parameters in the absence of turbulence are established. Hence, increasing the separation decreases the driving force for deposition, causing the deposit yield to decrease. This region is characterized by streamline particulate flow.
- The maxima of these data are consistent with the data shown in Figures 8 and 9. That is, the deposit yields are in the order 0.3 > 0.2 > 0.4 wt% PDADMAC for the standard conditions of time = 5 min time, voltage = 500 V, and electrode separation = 1 cm. Although these differences have been explained previously largely in terms of the coverage by the charging agent and the adhesive bonding by the polymer, it is possible that 0.2 wt% PDADMAC shows greater deposit yields than 0.4 wt% PDADMAC owing to additional effects possibly from reduced drag by the Ti particles with unsaturated surface coverage (viscosity is not considered relevant, as shown in Figure 3).
- It may be noted that the curves converge at an electric field of 200 V/cm (500 V/2.5 cm), which was noted previously as the minimal (threshold) electric field below which complete coverage for *initial* deposition was not observed. Examination of Figure 9 reveals that the first inflection for *continuing* deposition also occurs at the same electric field and at a deposit yield of ~3 mg/cm$^2$. If the electric field has been negated effectively at this point during deposition, then the data in Figure 10 confirm the threshold deposit thickness, which corresponds to a deposit yield of ~3 mg/cm$^2$.

## 4. Summary and Conclusions

- In the present work, two types of suspensions were used, these being the complete fraction (~1-50 μm; $d_{50}$ = ~17 μm) and a less sedimented fraction (~1-12 μm; $d_{50}$ = ~5 μm) of Ti in ethanol using PDADMAC as a charging agent. While there were no differences in the electrophoretic mobilities of these two types of suspensions, all subsequent data were obtained using only the complete fraction.
- The large particle size fraction (~20-50 μm) in these suspensions commenced sedimentation immediately following magnetic stirring and the intermediate particle size fraction (~12-20 μm) sedimented within 1 min, leaving suspended the fine particle size fraction (~1-12 μm).
- After 1 min, the less sedimented fraction of surface-charged particles of size ~1-12 μm acted as colloidal particles.
- PDADMAC levels in the range 0.05-0.3 wt% were found to be effective in reversing the surface charge of the Ti particles from negative to positive and the PDADMAC level of 0.3 wt% optimized their electrophoretic mobility. The apparent inflection in the electrical conductivity at this level supports this view.
- The electrophoretic mobility was used instead of the zeta potential to assess the response of the suspended Ti particles to the applied electric field. The former parameter allowed decoupling of the interactive effects embodied within the latter parameter, thereby permitting examination of the net effect of the charging agent on the Ti particles exclusive of the excess charging agent in solution.

- Using a simple algebraic method including the measurement of the electrophoretic mobility of the PDADMAC solutions, the electrophoretic mobility of Ti + Optimal PDADMAC could be decoupled from the electrophoretic mobility of Ti + Excess PDADMAC. By doing this, it was possible to differentiate the mechanistic stages of the process in terms of the PDADMAC level. This model was confirmed by the more time-consuming and laborious method of determining the deposit yield as a function of the PDADMAC level. Hence, this approach brings into focus a rapid and simple means of clarifying optimal additions of rheological aids.
- The three stages of the process are associated with the parameters of electrophoretic mobility and deposit yield as follows:

  (a) **Stage 1 –** An increase in both parameters resulted from the increasing adsorption of PDADMAC on the unsaturated surfaces of the Ti particles.

  (b) **Stage 2 –** Following saturation of the surfaces (at 0.3 wt% PDADMAC), there was a rapid decline in both parameters owing to increasing amounts of excess PDADMAC. This can be interpreted in terms of two scenarios: (i) sequential compression of the diffuse layer, reduction in the electric potential, decline in the electrophoretic mobility, and decrease in the deposit yield and (ii) concurrent competitive deposition of PDADMAC and Ti particles and consequent reduction in the deposition rate of the latter. The deposition of an insulating layer of polymer interposed between cathode and Ti particles reduced both the effect of the applied electric field and the adhesive strength of the deposition.

(c) **Stage 3** – The gradual levelling of both parameters (at ~1.5 wt% PDADMAC) also can be interpreted in terms of two scenarios: (i) Donnan potential constancy across the surface charge layer of soft particles, resulting in little or no decrease in electrophoretic mobility of optimally charged Ti particles with increasing ion concentration and (ii) establishment of a critical thickness of polymer-rich layer for the adhesion of the deposit, a critical electrical resistance on the cathode surface, and/or the dominance of the electrophoretic mobility of PDADMAC over the Ti particles.

- In the range of parameters studied, the deposit yield increased approximately logarithmically with increasing (a) solids loading, (b) deposition time, and (c) applied voltage. These observations are interpreted in light of Hamaker's equation for the deposit yield. The fourth major segment of the present work, the deposit yield as a function of (d) electrode separation, followed the Hamaker equation only in the streamline flow region but not the turbulent flow region.
- These four principal parameters were discussed in terms of the following considerations:

  (a) **Solids Loading** – The data are interpreted in terms of the decrease in electric field owing to the formation of what is effectively an electrically insulating layer of porous oxidized Ti particles and polymer.

  (b) **Deposition Time** – The data are interpreted in terms of the initial deposition of large and small particles (≤20 μm) close to the electrode during the first minute of deposition, followed over the next four minutes by deposition from a large reservoir of unsedimented smaller particles (~1-12 μm).

(c) **Applied Voltage** – The data are interpreted in terms of the low and high driving forces for deposition, corresponding to low and high voltages, respectively. More specifically, the deposit yield as a function of applied voltage is interpreted in terms of the presence of a threshold deposit yield, which caused the decrease or cessation of further deposition. These inflections were attributed to the establishment of a threshold electrically insulating layer of porous Ti and insulating polymer of a thickness sufficient to retard or stop further deposition.

(d) **Electrode Separation** – The data are interpreted in terms of the observed maxima, which represent the transition from turbulent to streamline flow in the suspensions. The curves of the data converge at an electric field of 200 V/cm, a value that apparently is confirmed by both the visual observations and the applied voltage data. That is, this electric field represents a minimal threshold to *initiate* deposition on the pristine cathode as well as to *continue* deposition of Ti following the establishment of the abovementioned electrically insulating layer. This layer was determined to correspond to a deposit yield of ~3 mg/cm$^2$.

## 5. Acknowledgements

The authors would like to thank the Universiti Teknikal Malaysia Melaka (UTeM) and the Ministry of Higher Education, Malaysia for financial support for this work.